\documentclass[10pt,aps,prb,reprint,superscriptaddress,amsfonts,amssymb,amsmath,floatfix,
longbibliography]{revtex4-2}
\usepackage{microtype}
\usepackage{lmodern}
\usepackage[T1]{fontenc}
\usepackage{graphicx}
\usepackage{hyperref}
\hypersetup{
    colorlinks=true,
    linkcolor=blue,
    citecolor=blue,   
    urlcolor=blue
}
\usepackage{epstopdf}%
\usepackage{xcolor}

\begin{document}

\title{Charge transfer between van der Waals coupled metallic 2D layers}

\author{Bharti Matta}
\email{b.matta@fkf.mpg.de}
\affiliation{Max-Planck-Institut für Festkörperforschung, Heisenbergstraße 1, 70569 Stuttgart, Germany}
\author{Philipp Rosenzweig}
\affiliation{Max-Planck-Institut für Festkörperforschung, Heisenbergstraße 1, 70569 Stuttgart, Germany}
\affiliation{Physikalisches Institut, Universität Stuttgart, Pfaffenwaldring 57, 70569 Stuttgart, Germany}
\author{Craig Polley}
\affiliation{MAX IV Laboratory, Lund University, Fotongatan 2, Lund 22484, Sweden}
\author{Ulrich Starke}
\author{Kathrin K{\"u}ster}
\email{k.kuester@fkf.mpg.de}
\affiliation{Max-Planck-Institut für Festkörperforschung, Heisenbergstraße 1, 70569 Stuttgart, Germany}
\date{\today}

\begin{abstract}

Van der Waals heterostructures have become a rapidly growing field in condensed matter research, offering a platform to engineer novel quantum systems by stacking different two-dimensional (2D) materials. A diverse range of material combinations, including hexagonal boron nitride, transition metal dichalcogenides and graphene, with electronic properties spanning from insulating to semiconducting, metallic, and semimetallic, have been explored to tune the properties of these heterostacks. However, understanding the interactions and charge transfer between the stacked layers remains challenging, particularly when more than two layers are involved. In this study, we investigate the charge transfer in a potassium-adlayer/graphene/lead-monolayer heterostructure stacked on a SiC substrate. Using synchrotron-based angle-resolved photoemission spectroscopy, we analyze the band structure of each layer, focusing on the charge transfer from K to the underlying 2D layers. Since K forms a {\mbox{(2${\times}$2)}} overlayer with respect to graphene, the amount of charge carriers donated by K can be determined. Our findings reveal that adsorption of K not only leads to a significant $n$-doping of the adjacent graphene layer but also to an electron transfer into the Pb monolayer. Remarkably, $\approx 44\%$ of the electrons donated by the K adlayer are transferred into its second nearest neighbouring layer, i.e. Pb, while $\approx 56\%$ remain in the graphene.

\end{abstract}

\maketitle

\section{Introduction} \label{sec1}
Two-dimensional (2D) materials have gained increasing attention since the exfoliation of monolayer graphene in 2004 \cite{NOVOSELOV2004}, owing to their unique electronic properties and potential for novel quantum phenomena \cite{BUTLER2013,MIRO2014}. A major breakthrough in this field is the controlled stacking of 2D materials to form van der Waals (vdW) heterostructures, unlocking entirely new and tunable physical properties \cite{GEIM2013,NOVOSELOV2016,GOMEZ2022}. Introducing a twist angle between layers further enriches their electronic behavior, as demonstrated in twisted bilayer graphene, where Moiré engineering enables correlated insulating states and superconductivity \cite{Cao2018, Cao2018correlated}. Despite weak vdW interactions, interlayer charge transfer and proximity coupling can lead to unconventional quantum systems \cite{NOVOSELOV2016,HU2022}. 
\\Intercalation of foreign atomic species at 2D interfaces provides another route for tailoring material properties \cite{YANG2024}. This method has been intensively used to modify the electronic structure of epitaxial graphene on SiC \cite{RIEDL2009,MCCHESNEY2010,EMTSEV2011,MARCHENKO2016,STOEHR2016,FORTI2016,LINK2019}, a leading platform for large-scale graphene growth on a technologically suitable substrate \cite{BERGER2004,EMTSEV2009,KRUSKOPF2016}. Intercalation of epitaxial graphene enables doping control from $p$-type \cite{WALTER2011,WOLFF2024} to nearly charge-neutral \cite{HAYASHI2017,MATTA2022}, all the way to strong $n$-doping, reaching its van Hove singularity \cite{LINK2019,ROSENZWEIG2019} and beyond \cite{ZAAROUR2023,HERRERA2024}. Apart from tuning graphene's properties, it is also a way to stabilize novel 2D forms of intercalant elements, exhibiting unique physical properties due to quantum confinement and inversion symmetry breaking \cite{FORTI2020,ROSENZWEIG2020,Gehring2025}.
\\Lead (Pb) has been extensively studied as an intercalant in recent years due to its superconducting nature down to the ultimate 2D limit \cite{Zhang2010} and strong spin-orbit coupling \cite{dil2008}. Previous studies have shown that intercalated Pb at the graphene/SiC interface shows a complex ordering with a very flat energy landscape, resulting in different superstructures, all closely related to a {\mbox{(1${\times}$1)}} order with respect to SiC \cite{YURTSEVER2016,CHEN2020,HU2021,GRUSCHWITZ2021,MATTA2022,otherHAN2023,SCHADLICH2023,VERA2024,SCHLOZEL2024,Matta2025}. Despite these structural variations, the system’s electronic structure seems to remain quite robust \cite{MATTA2022,VERA2024,SCHLOZEL2024,Matta2025}. Pb intercalation renders graphene nearly charge-neutral, with a residual hole density as low as $5.5\times10^{9}$ cm$^{-2}$ at room temperature \cite{MATTA2022}. The intercalated Pb layer exhibits a metallic band structure with a local band bottom at $\overline{\Gamma}$, dispersing towards the Fermi energy along the $\overline{\Gamma\mathrm{K}}$ and $\overline{\Gamma\mathrm{M}}$ directions, forming giant free-electron-like circular contours in the Fermi surface \cite{SCHADLICH2023,Matta2025}. 
\\Intercalated graphene heterostacks offer further tunability through adsorption of additional materials on the graphene surface. Alkali metal adsorption on epitaxial graphene is a well-established method for increasing the latter's electron density via surface charge transfer doping \cite{BOSTWICK2007,MCCHESNEY2010,WATCHARINYANON2011,WATCHARINYANON2012,ROSENZWEIG2020B,ROSENZWEIG2022,GOHLER2024}. This process can also transfer charge to the intercalant layer, potentially accessing previously unoccupied electronic states \cite{ROSENZWEIG2022,Gehring2025}. In this work, we deposit potassium (K) onto the graphene/Pb heterostack to induce electron transfer into the underlying layers. Using synchrotron-based angle-resolved photoemission spectroscopy (ARPES), we analyze the Fermi surfaces and band dispersion of the heterostack enabling us to quantify the charge transfer from K to both, the graphene and the Pb layer.  

\section{Experimental Details} \label{sec2}

Zerolayer graphene (ZLG) was grown on hydrogen-etched $n$-doped 6H-SiC(0001) wafer pieces (SiCrystal GmbH), as described elsewhere \cite{EMTSEV2009,FORTI2011,SOUBATCH2005}. Note that ZLG, which is the first hexagonally ordered carbon layer on SiC, remains partially covalently bonded to the substrate, saturating the silicon dangling bonds. As a result, the carbon atoms are not fully $sp^{2}$ hybridized, leading to significant deviations in the electronic structure of ZLG compared to a freestanding monolayer graphene (MLG) -- most notably, the absence of a Dirac cone. 
\\The ZLG samples were introduced into an ultra-high vacuum (UHV) system with a base pressure below $5\times10^{-10}$  mbar. After degassing the samples at $700$~$^{\circ}\mathrm{C}$ for 30 min, their quality was assessed using low-energy electron diffraction (LEED), confirming the characteristic $(6\sqrt{3}\times6\sqrt{3})\mathrm{R}30^\circ$ reconstruction. A $4$-$8$ nm thick Pb layer was then deposited onto the ZLG from a Knudsen cell, followed by stepwise annealing between $200$~$^{\circ}\mathrm{C}$ and $550$~$^{\circ}\mathrm{C}$ to intercalate the Pb atoms at the ZLG/SiC interface \cite{MATTA2022,Matta2025}. LEED was used to judge the intercalation quality: The $(6\sqrt{3}\times6\sqrt{3})\mathrm{R}30^\circ$ reconstruction spots were strongly suppressed, while the graphene diffraction spots became more intense than those of SiC. 
Furthermore, additional diffraction spots appeared around the first order graphene spots, corresponding to a {\mbox{(10${\times}$10)}} periodicity with respect to the graphene unit cell, indicating ordering in the Pb layer \cite{MATTA2022,SCHADLICH2023,Matta2025}. 
\\ARPES measurements were carried out at the Bloch beamline of the MAX IV synchrotron facility in Lund, Sweden. The Pb-intercalated quasi-freestanding monolayer graphene (Pb-QFMLG) samples were carried to the synchrotron in a UHV suitcase, maintaining a pressure below $1\times10^{-9}$ mbar during the transport. The samples were first degassed at $200$~$^{\circ}\mathrm{C}$ and their quality was verified using LEED, followed by ARPES measurements at $\approx 19$ K. For the sample investigated in detail here, a minor contribution from regions overgrown with MLG (not intercalated with Pb) was observed, which will be discussed in detail later. K was deposited \textit{in situ} onto the Pb-QFMLG sample at $\approx 19$ K from a commercial alkali-metal dispenser (SAES getters), operated at a current of $5.5$ A. Two deposition steps of about $30$ and $90$ seconds were carried out, after which the parabolic K 4s band and the newly formed $\pi$-band replicas became clearly visible (see Sec. \ref{sec3}).

\section{Results and Discussion} \label{sec3}

\begin{figure}[t]
	\centering
	\includegraphics{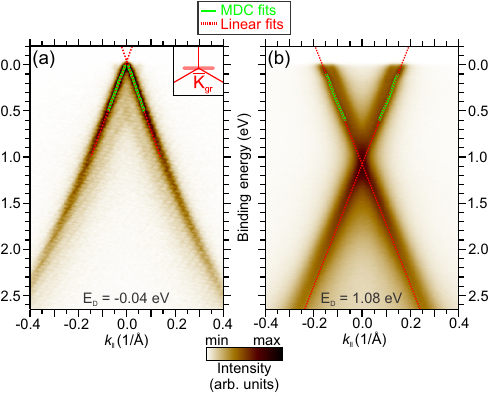}
	\caption{Dirac cone dispersion of Pb-QFMLG (a) before and (b) after K deposition: $E$-$k$ cuts at the $\overline{\mathrm{K}}$-point of graphene perpendicular to its $\overline{\Gamma\mathrm{K}}$ direction [see inset in (a)], measured at a photon energy of $40$ eV.} 
	\label{Fig_1}
\end{figure}

\begin{figure}[b]
	\centering
	\includegraphics{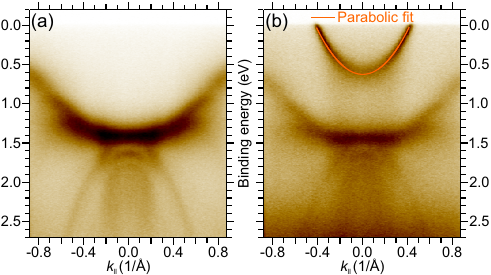}
	\caption{Pb bands along their $\overline{\mathrm{K}\Gamma\mathrm{K'}}$ direction, measured at $h\nu$ = $40$ eV, (a) before and (b) after K deposition. After K deposition, a band due to the K 4s electrons appears close to the Fermi energy, which is fitted with a parabola (orange) based on EDC and MDC fits.} 
	\label{Fig_2}
\end{figure}

Pb intercalation beneath ZLG results in nearly charge-neutral QFMLG, exhibiting only a negligibly small $p$-type doping \cite{MATTA2022,SCHADLICH2023}. However, the absolute charge carrier density in graphene depends on the sample temperature, effectively governed by the availability of free charge carriers in the $n$-doped SiC \cite{Matta2025}. Fig.~\ref{Fig_1}(a) shows the $\pi$-bands at the $\overline{\mathrm{K}}$-point of Pb-QFMLG, measured perpendicular to its $\overline{\Gamma\mathrm{K}}$ direction at $19$ K. Momentum distribution curve (MDC) fitting followed by linear extrapolation yields the Dirac point energy $E_D \approx 40$ meV above the Fermi energy, corresponding to a $p$-doping of $\approx 1.1\times10^{11}$ cm$^{-2}$. 
\\Subsequently, we deposited K onto Pb-QFMLG to both shift the respective electron bands to higher binding energies and investigate the charge distribution in the 2D layers. Fig.~\ref{Fig_1}(b) displays the $\pi$-bands after K deposition, clearly revealing a significant downshift of $E_D$ to $\approx 1.08$ eV below the Fermi energy. As mentioned, the pristine sample contained a minor amount of MLG, identifiable in Fig.~\ref{Fig_1}(a) by a second, faint Dirac cone with $E_D \approx 0.45$ eV \cite{RIEDL2010}. This MLG region, not intercalated with Pb, also undergoes electron doping from K, resulting in an apparent broadening of the graphene bands in Fig.~\ref{Fig_1}(b). A detailed analysis  for the MLG contribution yields $E_D \approx 0.83$ eV below the Fermi energy. After K deposition, the spectral intensity of MLG's $\pi$-bands increased relative to the $\pi$-bands of Pb-QFMLG. This could be due to more K atoms adhering to the Pb-QFMLG, thereby suppressing its spectral weight more than that of MLG. This is further supported by the smaller shift of $E_D$ for MLG ($\approx~0.4 $ eV) compared to Pb-QFMLG (more than 1.1 eV), indicating reduced charge transfer from K into the MLG -- likely due to fewer K atoms adsorbed on its surface. These observations suggest considerable mobility of K atoms on graphene, even when deposited at temperatures as low as $19$ K.
\\Fig.~\ref{Fig_2} presents the Pb bands measured along the $\overline{\mathrm{K}\Gamma\mathrm{K'}}$ direction of Pb, both before and after K deposition. In general, the intercalated 2D Pb layer exhibits a complex band structure that, at first glance, resembles a free-electron-like dispersion. However, features such as band splittings and avoided crossings, which are also evident from density functional theory calculations, are not fully captured by a simple free-electron approximation \cite{Matta2025}. Following K adsorption, the intensity of the Pb bands, as well as the SiC bands around $\overline\Gamma$, is noticeably suppressed. Additionally, an electron-like dispersion emerges around $\overline\Gamma$ close to the Fermi level, corresponding to the 4s valence band of the adsorbed K layer. MDC and energy distribution curve (EDC) fits confirm that this band follows a parabolic dispersion with an effective electron mass $m^* \approx 1.09~m_e$, comparable to K adsorption on Ag-intercalated QFMLG \cite{ROSENZWEIG2022}.
\begin{figure} [!h]
	\centering
	\includegraphics{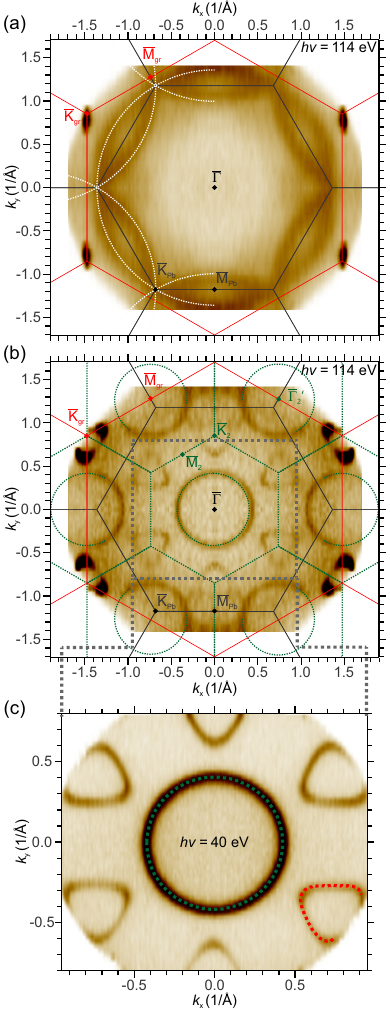}
	\caption{Fermi surface of Pb-QFMLG, (a) before and (b) after K deposition, measured at $h\nu$ = $114$ eV. The Pb and graphene BZs are indicated in black and red, respectively. The {\mbox{(2${\times}$2)}} superstructure of K with respect to graphene is indicated in green in (b). (c) Fermi surface after K deposition obtained at $h\nu$ = $40$ eV, covering the momentum space within the dotted gray box shown in (b). Note that the Fermi surfaces in (a) and (b) are four-fold symmetrized with respect to the $\overline{\mathrm{M}\Gamma\mathrm{M}}$ and $\overline{\mathrm{K}\Gamma\mathrm{K'}}$ directions of the Pb-BZ for visual enhancement.}
	\label{Fig_3}
\end{figure}
\\Figs.~\ref{Fig_3}(a) and (b) show the Fermi surface integrated up to $50$ meV below the Fermi energy, before and after K adsorption. The graphene and Pb Brillouin zones (BZs) are indicated by solid red and black hexagons, respectively. After K adsorption, the K 4s valence band appears as a circular contour around $\overline\Gamma$ (dashed green circle). The repeated circular contours, also highlighted with dashed green circles, suggest a {\mbox{(2${\times}$2)}} periodicity of the K adlayer relative to the graphene unit cell. This well-ordered K layer further supports the notion of significant adatom mobility on the surface. Additionally, graphene $\pi$-band replicas emerge at the $\overline{\mathrm{K}}_2$ points of the respective {\mbox{(2${\times}$2)}} BZ, originating from a translational displacement of the primary $\overline{\mathrm{K}}$ points along the {\mbox{(2${\times}$2)}} reciprocal lattice vectors. Such {\mbox{(2${\times}$2)}} ordering is well known for alkali metals adsorbed on bulk graphite \cite{WU1982,HU1986,BENNICH1999,BREITHOLTZ2004} and has also been observed for K adsorption on Ag-intercalated QFMLG \cite{ROSENZWEIG2022}. In contrast, for K adsorption on highly doped Yb-intercalated graphene \cite{ROSENZWEIG2020B} and bismuthene-intercalated graphene \cite{Gehring2025}, no dispersive K 4s band was observed near $\overline{\Gamma}$. Note that the $\pi$-band replicas shown in the high-resolution Fermi surface map in Fig.~\ref{Fig_3}(c) only consist of one Dirac cone stemming from the K-doped Pb-QFMLG and the second Dirac cone from MLG is not visible. This indicates that on the MLG, no {\mbox{(2${\times}$2)}} ordering of K is present, probably due to the smaller amount of K adatoms adsorbed on MLG as discussed earlier.
\begin{figure*}[t]
	\centering
	\includegraphics{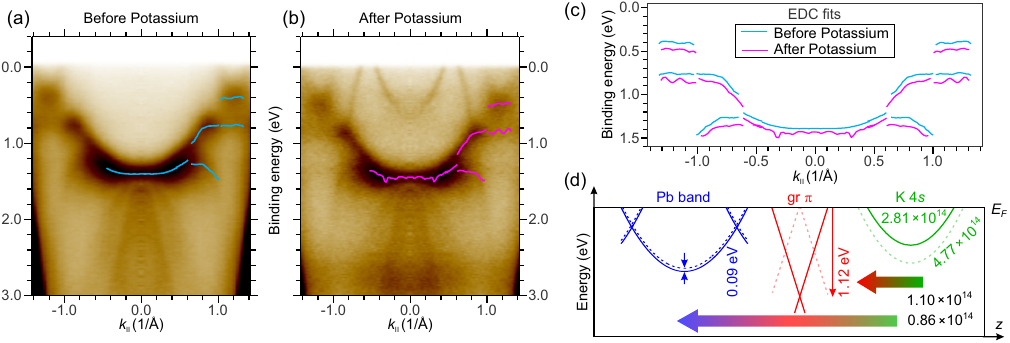}
	\caption{$E-k$ dispersion along the $\overline{\mathrm{M}\Gamma\mathrm{M}}$ direction of Pb, (a) before and (b) after K deposition, respectively measured at $h\nu$ = $114$ eV. EDC fits through the Pb band dispersion (blue and pink curves) show a clear downshift of the bands after K adsorption, as demonstrated in (c). (d) Schematic drawing of the charge transfer between K, graphene and Pb (not drawn to scale). The dashed (solid) curves indicate the band position before (after) K adsorption. Charge carrier concentrations are given in cm$^{-2}$.}
	\label{Fig_4}
\end{figure*}
\\Based on the fit of the K 4s valence band displayed by the green circle in Fig.~\ref{Fig_3} (c), an average Fermi momentum $k_F$ of $0.42\pm0.01~\text{\AA}^{-1}$ was derived, corresponding to a carrier density of $(2.81\pm0.13)\times10^{14}$ cm$^{-2}$ according to Luttinger's theorem \cite{LUTTINGER1960,LUTTINGER1960a}. Assuming that each {\mbox{(2${\times}$2)}} unit cell of the K adlayer hosts one K atom and consequently, one 4s electron, the total valence electron density of the K adlayer should be $4.77\times10^{14}$ cm$^{-2}$ \cite{ROSENZWEIG2022}. This implies that the difference of $(1.96\pm0.13)\times10^{14}$ cm$^{-2}$ is donated into the graphene/Pb heterostack. Furthermore, based on the average area of the replicated $\pi$-pockets in Figs.~\ref{Fig_3} (b) and (c) [red dashed fit in Fig.~\ref{Fig_3}(c)], the charge carrier density of graphene is determined to be $(1.10\pm0.03)\times10^{14}$ cm$^{-2}$. Note that the pristine Pb-QFMLG before K deposition exhibited a hole concentration of $\approx 1.1\times10^{11}$ cm$^{-2}$. The amount of electrons in graphene after K deposition accounts for only {$\approx56\%$} of the charge transferred from the {\mbox{(2${\times}$2)}} K adlayer, suggesting that the remaining electron density of $(0.86\pm0.13)\times10^{14}$ cm$^{-2}$ either goes to the Pb layer or to some unsaturated dangling bonds of SiC, or both. 
\\Next, we analyze the electron density in the interlayer Pb. For pristine Pb-QFMLG, the free-electron-like contours corresponding to the first and repeated BZs of the Pb interlayer are clearly visible in Fig.~\ref{Fig_3}(a). This dispersion can be described using a free-electron approximation (white dashed curves) with $k_F = 1.36\pm0.01~\text{\AA}^{-1}$, $m^* = 5.0~m_e$ and $a = a_{SiC} = 3.08~\text{\AA}$ (lattice constant) \cite{SCHADLICH2023, Matta2025}. The $k_F$ value was derived from MDCs at the Fermi energy along the $\overline{\mathrm{M}\Gamma\mathrm{M}}$ direction of Pb. The corresponding electron density was obtained by applying Luttinger's theorem \cite{LUTTINGER1960,LUTTINGER1960a} to the Fermi surface area in the first BZ of Pb, yielding $(5.09\pm0.43)\times10^{14}$ cm$^{-2}$. Note that this value is slightly lower than the Pb electron density published earlier \cite{Matta2025}. We relate this to small deviations in the amount of Pb atoms intercalated. In our previous research, we have shown that the Pb atoms form a slightly disordered grain boundary network \cite{SCHADLICH2023} and depending on the defect density of the pristine ZLG and the exact intercalation procedure, it is likely that the amount of Pb atoms in the grain boundaries varies slightly, which can consequently affect the electron density of the Pb layer.
\\In order to determine changes in the electron density of the interlayer Pb after K deposition, it is essential to analyze how the Pb bands evolve in the Fermi surface. However, their intensity nearly vanishes close to the Fermi energy due to the overall quenching of their spectral weight after K deposition. Additionally, they are overshadowed by the intense K 4s bands in the repeated BZs of K [dashed green circles in Fig.~\ref{Fig_3}(b)] and by the replicated $\pi$-bands, which appear in the momentum range where the Pb bands are expected to cross the Fermi surface in their first BZ. This renders a precise determination of the Pb carrier concentration based on the Fermi surface impossible.
\\Insights into the electronic structure of the Pb layer after K adsorption can also be gained from the band dispersion further away from the Fermi energy. Therefore, we have performed EDC fits for the Pb band dispersion along the $\overline{\mathrm{M}\Gamma\mathrm{M}}$ direction, before and after K adsorption [Figs.~\ref{Fig_4}(a)-(c)]. After K adsorption, the Pb bands shift towards higher binding energy. The energy shift as a function of $k_{||}$ varies from about $0.04$ to $0.14$ eV. Note that the band bottom of Pb around the $\overline{\Gamma}$ point shifts the least in energy, which could be due to its hybridization with the SiC valence bands \cite{MATTA2022, Matta2025}. The observation of slightly varying shifts of the different band features might indicate some small renormalizations in the Pb bands. Based on the energy shifts, we performed a rigid shift of the free-electron approximation derived from the Pb bands before K deposition by $0.09\pm0.05$ eV\footnote{The energy shift value was obtained by averaging the minimum and maximum observed shifts of $0.04$ and $0.14$ eV, and the error bar represents half their difference.} in order to obtain an estimate for the charge transferred into Pb. The corresponding increase in $k_F$ is $0.04\pm0.02~\text{\AA}^{-1}$ compared to the pristine Pb-QFMLG, giving an electron density of $(6.85\pm0.89)\times10^{14}$ cm$^{-2}$ in the interlayer Pb after K adsorption. Based on this estimate, the increase in the electron density of the Pb layer after K deposition is $(1.76\pm0.99)\times10^{14}$ cm$^{-2}$. The expected value of $(0.86\pm0.13)\times10^{14}$ cm$^{-2}$ calculated from the electron density that neither remains in the K adlayer, nor is transferred into graphene, falls within the uncertainty bounds of the estimated value. Therefore, we can conclude that this charge difference is transferred into the Pb layer. In Fig. ~\ref{Fig_4}(d), the energy band shifts are sketched, and the amount and direction of the charges transferred between K, graphene and the Pb layer are indicated.
\\It is quite remarkable that from the charge donated by K, $\approx56\%$ remains in graphene while $\approx44\%$ is further transferred into the second nearest neighbouring layer (i.e., the Pb layer). However, the fact that the pristine Pb-intercalated graphene exhibits a slight $p$-doping, despite its proximity to the Pb layer with a high electron density of $\approx5.09\times10^{14}$ cm$^{-2}$, indicates that electron transfer from Pb into graphene is energetically unfavorable, and Pb most-likely acts as electron acceptor rather than electron donor. 
\\Finally, we compare the charge transfer observed for Pb-QFMLG with that in the Ag-QFMLG case \cite{ROSENZWEIG2022}. For K-doped Ag-intercalated graphene, both the carrier density remaining in K and the one donated to graphene are slightly larger than in the present case of Pb-QFMLG. Hence, only $0.55\times10^{14}$ cm$^{-2}$ electrons neither remain in K nor are transferred into graphene for Ag-QFMLG. Since the intercalated Ag is a semiconducting layer and the charge transfer is not large enough to fill its conduction band, it is speculated that this residual carrier density is transferred to the dangling bonds of SiC. In fact, Fermi level pinning within the semiconducting gap was observed and the Ag valence band was found to shift rigidly with the Si $2p$ core level states by 0.26 eV. In contrast, Pb-QFMLG exhibits a small interlayer band shift on the order of 90 meV with no observable shifts of the Si core levels. We conclude that for Pb-QFMLG the residual charge carriers (i.e., the ones missing in K that are not found in graphene) are not saturating dangling bonds of Si but are instead transferred into the Pb layer. This may be attributed to the dense packing of the Pb atoms in the grain boundary structure, which likely results in the prior saturation of most Si dangling bonds, and to the availability of unoccupied states just above the Fermi energy for the metallic Pb layer. 

\section*{Conclusions}

We have investigated the influence of K adsorption on Pb-intercalated quasi-freestanding monolayer graphene on SiC. The K adlayer forms a {\mbox{(2${\times}$2)}} superstructure relative to the graphene unit cell, giving rise to a parabolic electron band of K 4s electrons and $\pi$-band replicas appearing at the $\overline{\mathrm{K}}_2$ points of the corresponding {\mbox{(2${\times}$2)}} Brillouin zone. K deposition induces significant $n$-doping in graphene ($\approx 1.1\times10^{14}$ cm$^{-2}$), as opposed to a slight $p$-doping ($\approx 1.1\times10^{11}$ cm$^{-2}$) in the pristine sample. Based on the {\mbox{(2${\times}$2)}} superstructure, we estimate the expected electron density in the ordered K layer. By analyzing the charge remaining in the K 4s band and the amount transferred to graphene, we infer that an electron density of $\approx 0.9\times10^{14}$ cm$^{-2}$ is transferred into the Pb monolayer, i.e., the second nearest neighboring layer of K. This charge transfer could be qualitatively confirmed by a shift of the Pb bands away from the Fermi energy and a corresponding increase of $0.04\pm0.02~\text{\AA}^{-1}$ in the Fermi wave vector of their quasi-free-electron approximation. It is remarkable that $\approx 44\%$ of the charges transferred from K go into the more distant Pb layer while $\approx 56\%$ remain in the adjacent graphene.
\\This study demonstrates that interlayer charge transfer in van der Waals heterostacks extends beyond adjacent layers, influencing more distant ones as well. Even metallic layers with inherently high electron densities (Pb in this case) can undergo further electron doping. This is in contrast with the case of Ag, a semiconducting interlayer, where the K adlayer induces a comparable charge transfer into graphene but results in a significantly larger energy shift of the interlayer bands, despite a smaller charge transfer compared to Pb. The possibility to tune the charge carrier concentration in a Pb layer confined underneath graphene may also be of interest for potential applications in the field of electronics and spintronics. 

\begin{acknowledgments}
This work was supported by the Deutsche Forschungsgemeinschaft (DFG, German Research Foundation) within 
the FLAG-ERA framework through Project Sta315/9-1 and within the Research Unit FOR5242 through Projects 
Ku4228/1-1 and Sta315/13-1. MAX IV Laboratory is acknowledged for time on beamline BLOCH under proposal 20210067. We are also thankful to the entire beamline staff at BLOCH for their great support. Research conducted at MAX IV, a Swedish national user facility, is supported by the Swedish Research council under contract 2018-07152, the Swedish Governmental Agency for Innovation Systems under contract 2018-04969, and Formas under contract 2019-02496. 
\end{acknowledgments}

\end{document}